\documentclass[aps,prl,reprint,superscriptaddress,notitlepage]{revtex4-1}
\usepackage{times,amsmath,amsfonts,amssymb,mathrsfs,graphics,graphicx,color,comment,bm}
\usepackage[next]{inputenc}
\usepackage[dvips]{epsfig}
\usepackage{hyperref}
\usepackage{pdfsync}
\usepackage{appendix}
\usepackage{textcomp}
\usepackage{lipsum}  
\usepackage{gensymb}

\begin{document}

\title{Stability and Metastability of Skyrmions in Thin Lamellae of Cu$_2$OSeO$_3$}

\author{M. N. Wilson}
\address{Durham University, Department of Physics, South Road, Durham, DH1 3LE, United Kingdom}
\author{M. T. Birch}
\address{Durham University, Department of Physics, South Road, Durham, DH1 3LE, United Kingdom}
\address{Diamond Light Source, Didcot, OX11 0DE, United Kingdom}
\author{A. {\v S}tefan{\v c}i{\v c}}
\address{University of Warwick, Department of Physics, Coventry, CV4 7AL, United Kingdom}
\author{A. C. Twitchett-Harrison}
\address{University of Cambridge, Department of Materials Science and Metallurgy, Cambridge, CB3 0FS, United Kingdom}
\author{G. Balakrishnan}
\address{University of Warwick, Department of Physics, Coventry, CV4 7AL, United Kingdom}
\author{T. J. Hicken}
\address{Durham University, Department of Physics, South Road, Durham, DH1 3LE, United Kingdom}
\author{R. Fan}
\address{Diamond Light Source, Didcot, OX11 0DE, United Kingdom}
\author{P. Steadman}
\address{Diamond Light Source, Didcot, OX11 0DE, United Kingdom}
\author{P. D. Hatton}
\address{Durham University, Department of Physics, South Road, Durham, DH1 3LE, United Kingdom}

\begin{abstract}
We report small angle X-ray scattering (SAXS) measurements of the skyrmion lattice in two 200~nm thick Cu$_2$OSeO$_3$ lamellae aligned with the applied magnetic field parallel to the out of plane [110] or [100] crystallographic directions. Our measurements show that the equilibrium skyrmion phase in both samples is expanded significantly compared to bulk crystals, existing between approximately 30 and 50~K over a wide region of magnetic field. This skyrmion state is elliptically distorted at low fields for the [110] sample, and symmetric for the [100] sample, possibly due to crystalline anisotropy becoming more important at this sample thickness than it is in bulk samples. Furthermore, we find that a metastable skyrmion state can be observed at low temperature by field cooling through the equilibrium skyrmion pocket in both samples. In contrast to the behavior in bulk samples, the volume fraction of metastable skyrmions does not significantly depend on cooling rate. We show that a possible explanation for this is the change in the lowest temperature of the skyrmion state in this lamellae compared to bulk, without requiring different energetics of the skyrmion state.

\end{abstract}
\maketitle

\section{Introduction}

Recent research in condensed matter physics has put considerable emphasis on topology and topological magnetism~\cite{Duncan2017,Hasan2010,Qi2012,Nagaosa2013}. The magnetic skyrmion lattice is one topological state of matter that has attracted significant interest~\cite{Rossler2006}. It typically consists of a hexagonal array of topologically protected magnetic vortex-like structures that appear in a variety of different materials, usually stabilized by a combination of symmetric exchange, the Dzyaloshinskii-Moriya interaction (DMI), crystal anisotropies, and thermal fluctuations~\cite{Nagaosa2013}. Since their first discovery in the B20 metal MnSi~\cite{Muhlbauer2009}, skyrmions have been found in similar noncentrosymmetric materials such as FeGe~\cite{Yu2011}, Fe$_{1-x}$Co$_x$Si~\cite{Munzer2010}, Cu$_2$OSeO$_3$~\cite{Seki2012}, and others~\cite{Tokunga2015, Fujima2017,Ruff2015}, and recently in some centrosymmetric materials where geometric magnetic frustration is thought to play a role~\cite{Kurumaji2018, Hirschberger2018}. They have also been seen in grown thin films and multilayers where interfacial DMI~\cite{Heinze2011, Moreau2016, Anjan2017} or a combination of DMI, uniaxial anisotropy, and geometric confinement  \cite{Leonov2016,Wilson2014,Rybakov2013} help stabilize the skyrmions.

While skyrmions in bulk are typically only stable near the critical temperature ($T_C$), it has been found in a number of materials that a metastable skyrmion state can be formed by rapidly field cooling the system through the equilibrium skyrmion pocket~\cite{Okamura2016,Milde2013,Oike2016,Karube2016}. This state is not the energy minimum of the system, however the energy barrier to transition out of it is large enough that at low temperatures no decay of the metastable skyrmion state is observed in experiments; they persist for periods longer than a week \cite{Oike2016}. Nevertheless, near the lower boundary of the equilibrium skyrmion state, the lifetime is observed to become very short (seconds)~\cite{Oike2016, Wilson2019}. As a result, large cooling rates are required to stabilize a significant population of metastable skyrmions in bulk samples ($> 40$~K/min for 50\% skyrmion population in bulk Cu$_2$OSeO$_3$) \cite{Birch2019}. This presents a limitation to the study and use of metastable skyrmions, which has been previously alleviated using rapid heat pulses from lasers \cite{Berruto2018} or electrical heaters \cite{Bannenberg2019} to form the metastable state, both of which are very energy intensive methods. 

In a number of materials it has been observed that the equilibrium skyrmion region is stable over a significantly wider temperature range in thin films than in bulk crystals \cite{Yu2011, Seki2012, Morikawa2017, Stolt2019, Yu2015}. One intriguing example of these materials is Cu$_2$OSeO$_3$, which is a multiferroic insulator that hosts magnetic skyrmions in a phase diagram that is qualitatively very similar to that of MnSi. The particular interest in this material comes because the skyrmions carry an electric polarization \cite{Seki2012a}, and their stability can be manipulated by the application of electric fields in bulk crystals \cite{Kruchkov2018} or thin lamellae \cite{Huang2018}. 

Studies of thin lamellae of Cu$_2$OSeO$_3$ with Lorentz transmission electron microscopy (LTEM) show that skyrmions are stable over a very wide temperature range, from $T_C = 58$~K down to at least 5~K \cite{Seki2012}, however this and other LTEM work was necessarily done with a very thin lamella (approximately 100~nm or less) to allow electron transmission. Work has also been recently been performed on somewhat thicker lamellae ($> 200$~nm) of Cu$_2$OSeO$_3$ with small angle X-ray scattering (SAXS), studying the effect of electric fields and strain in one orientation, but not looking in detail at the metastable state \cite{Okamura2017, Okamura2017E}. More information on the behavior of the equilibrium and metastable skyrmion states in thin lamellae of Cu$_2$OSeO$_3$ will be important to inform future research into skyrmions. In particular, understanding how they are affected by crystal orientation is important, as this has been found to be significant in bulk samples \cite{Halder2018}.
 
In this paper, we use small angle X-ray scattering to probe the skyrmion state in $200 \pm 25$~nm thick plates of Cu$_2$OSeO$_3$  with the field and the film normal parallel to the [110] or [100] crystal axis. We find that for both orientations, the equilibrium skyrmion stability region is dramatically expanded compared to bulk samples, similar to previous reports. The phase diagram for the two orientations are very similar, however the [110] sample shows an elliptical distortion in the skyrmion state at low field, while the [100] sample does not. Our data also shows that metastable skyrmions can be observed at low temperature by field cooling through the equilibrium skyrmion state. In contrast to the behavior in bulk Cu$_2$OSeO$_3$, the volume fraction of this metastable state does not strongly depend on cooling rate, indicating that the decay rate of metastable skyrmions is much slower in thin samples.

\section{Experiment Details}

A single crystal of Cu$_2$OSeO$_3$ was grown from pre-reacted polycrystalline powder using chemical vapour transport with TeCl$_4$ as a transporting agent (Details can be found in the supplemental information of Ref. \cite{Stefancic2018}). An FEI Helios nano lab dual-beam was used to prepare $200 \pm 25$ nm thick lamellae with lateral dimensions of 4 x 6 $\mu$m in [100] and [110] orientations. Thickness and sample size was determined using scanning electron microscopy inside the Helios nanolab. The lamellae were lifted out using an Omniprobe micromanipulator, placed on 200 nm thick silicon nitride membranes, and attached using an ion-beam deposited platinum strap. The membranes were produced by Silson Ltd. with nine 0.5 x 0.5~mm Si$_3$N$_4$ windows on a 5 x 5~mm 200~$\mu$m thick silicon frame. Prior to attachment of the lamellae, we sputter-coated the membranes with 600~nm of gold and ion-milled 3~$\mu$m apertures in the gold, over which the samples were placed. This gold coating is sufficient to stop the vast majority of soft X-rays incident on the membrane, minimizing background from X-rays that miss the sample. 

The silicon frame with attached lamellae was then mounted on a copper sample holder on the cryostat cold finger of the RASOR instrument \cite{Beale2010} at the Diamond Light Source. Cooling to a base temperature of 20~K was achieved using a liquid helium cold finger cryostat. Magnetic fields were applied to the sample using a four-pillar permanent magnet array, which was moved relative to the sample to change the applied field, and calibrated using a Hall probe. Using this instrument, we performed small angle resonant soft X-ray scattering measurements using circularly polarized X-rays. These measurements were performed with the sample in transmission geometry, and X-rays were detected using a CCD camera with 2048 $\times$ 2048 13.5~$\mu$m pixels mounted 138~mm from the sample, giving an effective $Q$ resolution of 1 x 10$^{-3}$~nm$^{-1}$ at the copper $L_3$ absorption edge ($2p$ to $3d$ transition). A horizontal beam stop was used to block the main transmitted beam to avoid damaging the detector. A schematic diagram of the scattering experimental setup is shown in Fig. \ref{fig:Energy} (a).

\section{Results and Discussion}

\begin{figure}[t]
\includegraphics[width=\columnwidth]{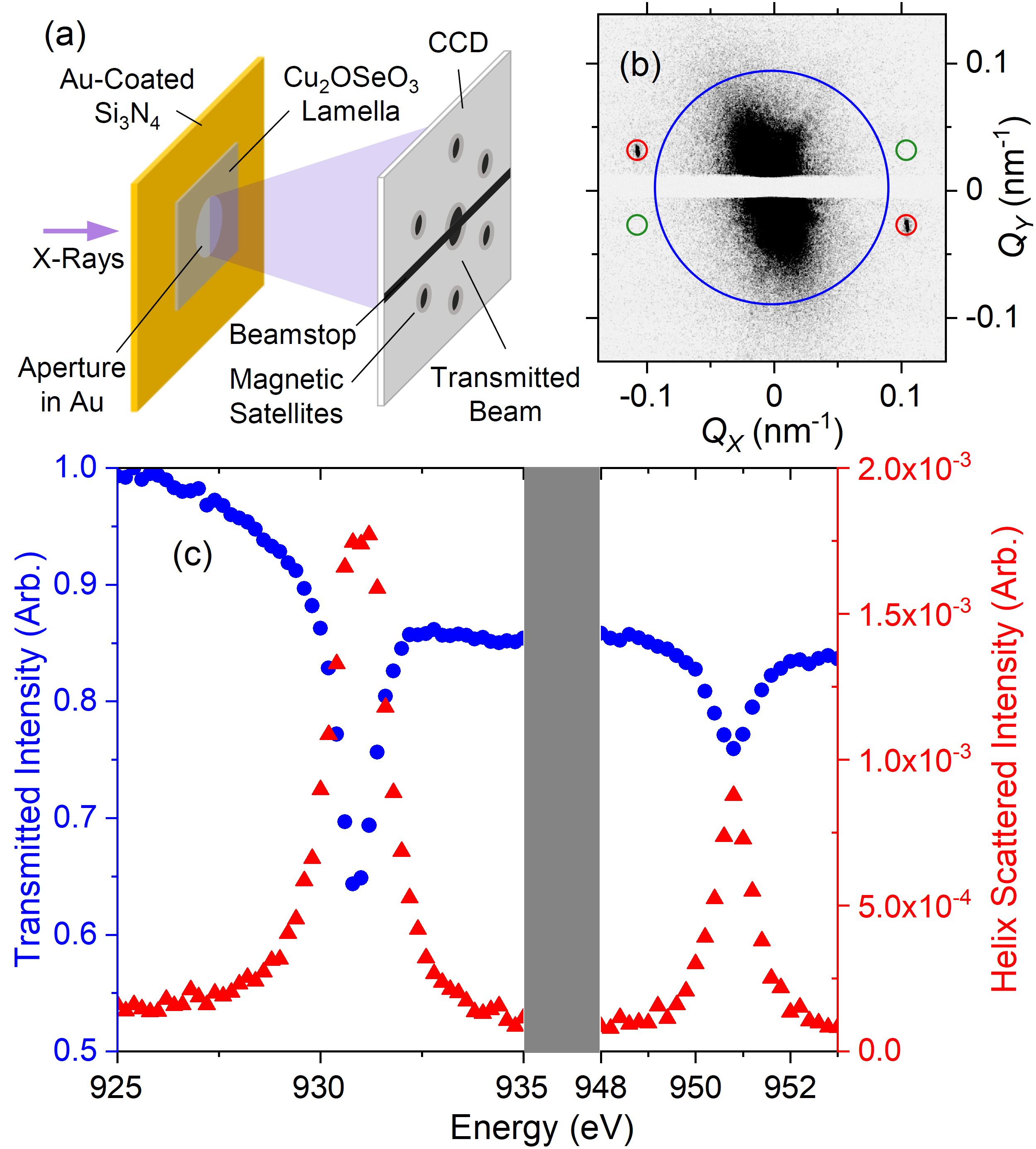}
\caption{(a) Schematic diagram of the experimental setup. (b) Small angle X-ray scattering pattern of the helical state of Cu$_2$OSeO$_3$. (c) Energy scan of X-ray transmission (blue) and magnetic scattering (red) for a 200 nm thick Cu$_2$OSeO$_3$ lamella at 20~K and zero magnetic field, aligned with the incoming beam $\parallel$ [110]. Representative transmitted intensity was determined by summing inside the blue circle on (b), and the scattered intensity was taken by summing the red circles and subtracting the green circles as background.}
\label{fig:Energy}
\end{figure}

Figure \ref{fig:Energy} (b) shows a SAXS pattern of the [110] oriented Cu$_2$OSeO$_3$ sample taken in zero magnetic field at 20~K. In this image, the  peaks at approximately $Q_X = +/-  0.1$~nm$^{-1}$ arise from  magnetic scattering from the helical state of the sample, corresponding to a helical period of $56 \pm 0.5$~nm, while the intensity in the center corresponds to over-spill of the transmitted beam around the beam stop as well as some coherent charge scattering from the aperture geometry. In all other SAXS patterns shown in this paper, the central charge scattering feature was minimized by subtracting background images taken where no magnetic scattering appeared, and noise in the images was reduced by applying a 5 x 5 pixel Wiener filter to the images. To determine the optimal energy for magnetic measurements, we characterized the magnetic scattered intensity as a function of energy by summing the helical scattering peak (indicated by the red circles on Fig. \ref{fig:Energy} (b)), subtracting the symmetry-equivalent green circles as background, and used the sum of the blue circle of Fig. \ref{fig:Energy} (b) as a proxy for the transmitted beam intensity. The result of these measurements is shown in Fig. \ref{fig:Energy} (c), where the intensities have been scaled so that a value of 1 corresponds to the peak transmission. We see two clear minima in the transmitted intensity, and associated maxima in the magnetic scattering, at energies of 931~eV and 951~eV, which match the expected energies of the $L_3$ and $L_2$ absorption edges of copper. For the remainder of the measurements reported in this paper we used an X-ray energy of 931~eV where we measured maximum magnetic scattering.

\begin{figure}[t]
\includegraphics[width=\columnwidth]{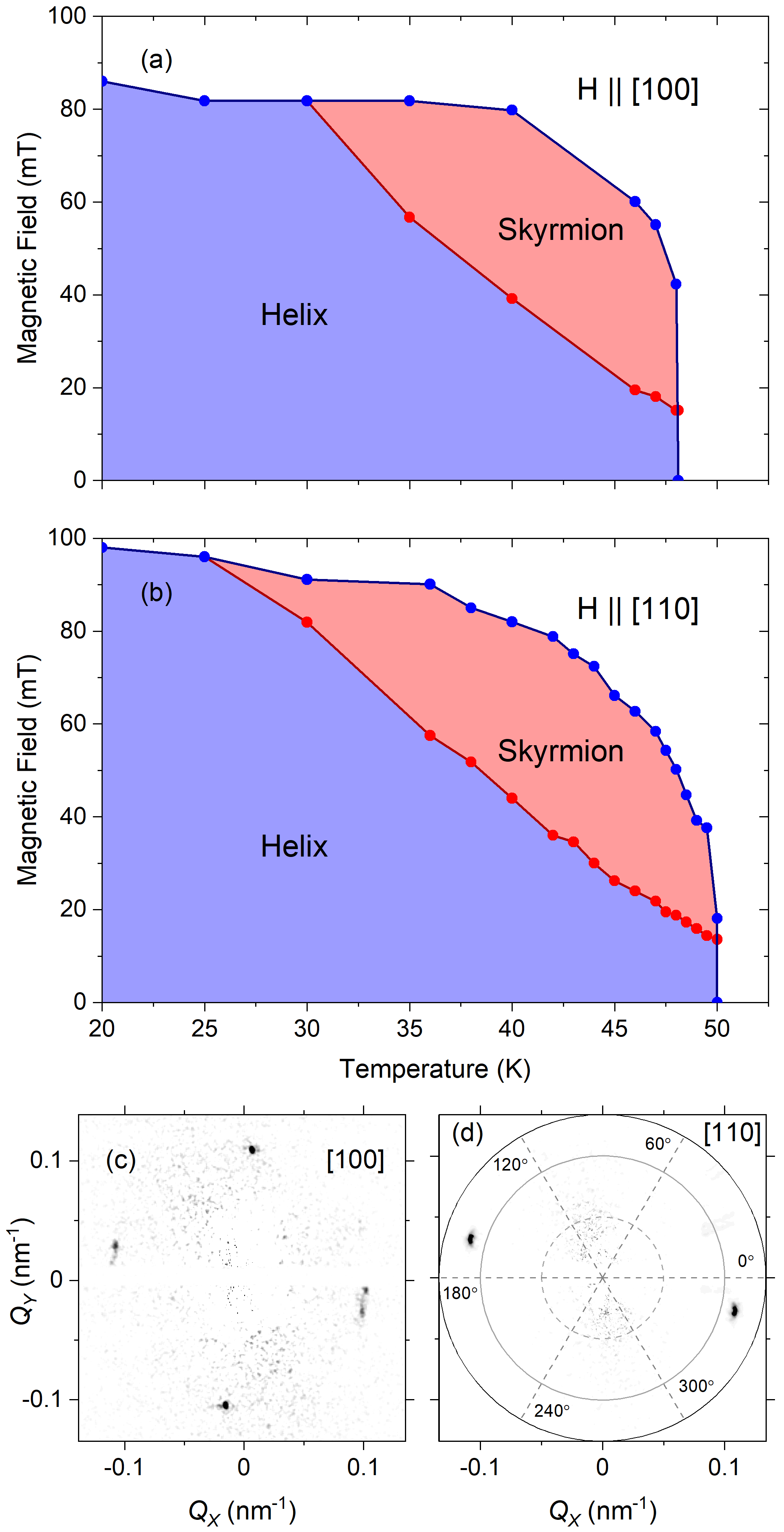}
\caption{Phase diagram of the [100] (a) and [110] (b) samples on zero field cooling, measuring with increasing field. Red shows the region where a skyrmion lattice appears and, blue shows the region where a helical state appears, and white shows where no magnetic scattering occurs, indicating either a field polarized, conical, or paramagnetic state.  Panels (c) and (d) show the characteristic helical state in both samples measured at 6~mT and 20~K. The angular convention shown in (d) is used to label peaks in the remainder of the paper.}
\label{fig:Phase}
\end{figure}

Using SAXS, we then investigated the magnetic phase diagram of the two samples as a function of temperature and magnetic field by performing zero field cooled magnetic field sweeps at various temperatures between 20 and 60~K. The result of these measurements is shown in Fig. \ref{fig:Phase} (a) for the [100] sample, and Fig. \ref{fig:Phase} (b) for the [110] sample, where blue represents the region where a helical state was seen, red the skyrmion state, and white where no magnetic scattering was observed, corresponding likely to either the field polarized or paramagnetic states. These diagrams show little qualitative difference between the two samples, with a skyrmion state existing for both in a wide range of magnetic field between 30 and 50~K, and a helical state existing up to above 80~mT for both samples.The upper magnetic field is slightly different for the two, which may be due to a combination of small differences in demagnetizing fields on the two samples from slightly different sample shapes, and magnetocrystalline anisotropy. One significant difference in the behavior is, however, demonstrated in figures \ref{fig:Phase} (c) and (d), which show representative helical SAXS patterns for the [100] and [110] samples measured at 6~mT and 20~K. For the [110] sample, as previously mentioned, two symmetry-equivalent peaks are observed for the helical state, corresponding to a single helical domain. However, for the [100] sample, we see two pairs of peaks, which corresponds to two separate helical domains existing in the sample. This can be understood from the preference for bulk Cu$_2$OSeO$_3$ to form helices at low field along any of the $\langle100\rangle$ family directions. In the [110] sample there is only one of these directions existing in plane, resulting in a single helical domain. However, in the [100] sample, there are two, resulting in the two helical domains we see appearing approximately 90\degree ~apart. 

These phase diagrams contrast dramatically with the phase diagram seen for bulk Cu$_2$OSeO$_3$, where the helical state transitions to the conical state (which would not show scattering peaks in our SAXS geometry)  by 25~mT [110] or 100~mT [100], and the skyrmion state only extends over a small field range between 56 and 58~K \cite{Halder2018}. These differences likely arise primarily from anisotropies caused by the thinning of the samples making it unfavorable for the helical propagation vector to be directed out of plane. This would make the conical state energetically unfavorable, allowing the helical state to exist up to higher magnetic fields and expanding the stability region of the skyrmion pocket. The difference in critical temperature between what we observe and the bulk sample measurements possibly arises from a small temperature gradient between the thermometer and the samples, which sit on a thin silicon nitride membrane that has imperfect thermal contact with the cryostat. This would make the sample temperature slightly higher than the recorded temperature at the thermometer position. There may also be a slight additional suppression in the transition temperature in the [100] sample due to differing small amounts of ion damage from the process of producing a lamella using the focused ion beam, however we expect this to be minimal as the sample preparation method was identical for both samples.

 Broadly, our phase diagram shows some similarity to other reports of the magnetic behavior of nominally 200~nm thick plates of [110] oriented Cu$_2$OSeO$_3$ by SAXS \cite{Okamura2017, Okamura2017E}, however we find skyrmions to exist over a wider temperature range than the 10~K seen by Okamura et al. \cite{Okamura2017}. The size of our measured stability region falls between this report and that seen in 100~nm thick plates by LTEM \cite{Seki2012}, likely suggesting that our 200~nm thick films are somewhat thinner than those that Okamura et al. report to be ``about 200-nm'' \cite{Okamura2017}. The additional measurements we make on a [100] oriented sample further show that, similar to what is seen in bulk samples above 20~K \cite{Halder2018}, the overall magnetic phase diagram is not strongly modified by crystal direction in thinned samples. This suggests that in our thinned samples, shape anisotropy dominates over crystalline anisotropy in determining the stability regions of the magnetic phases.

\begin{figure}[]
\includegraphics[width=\columnwidth]{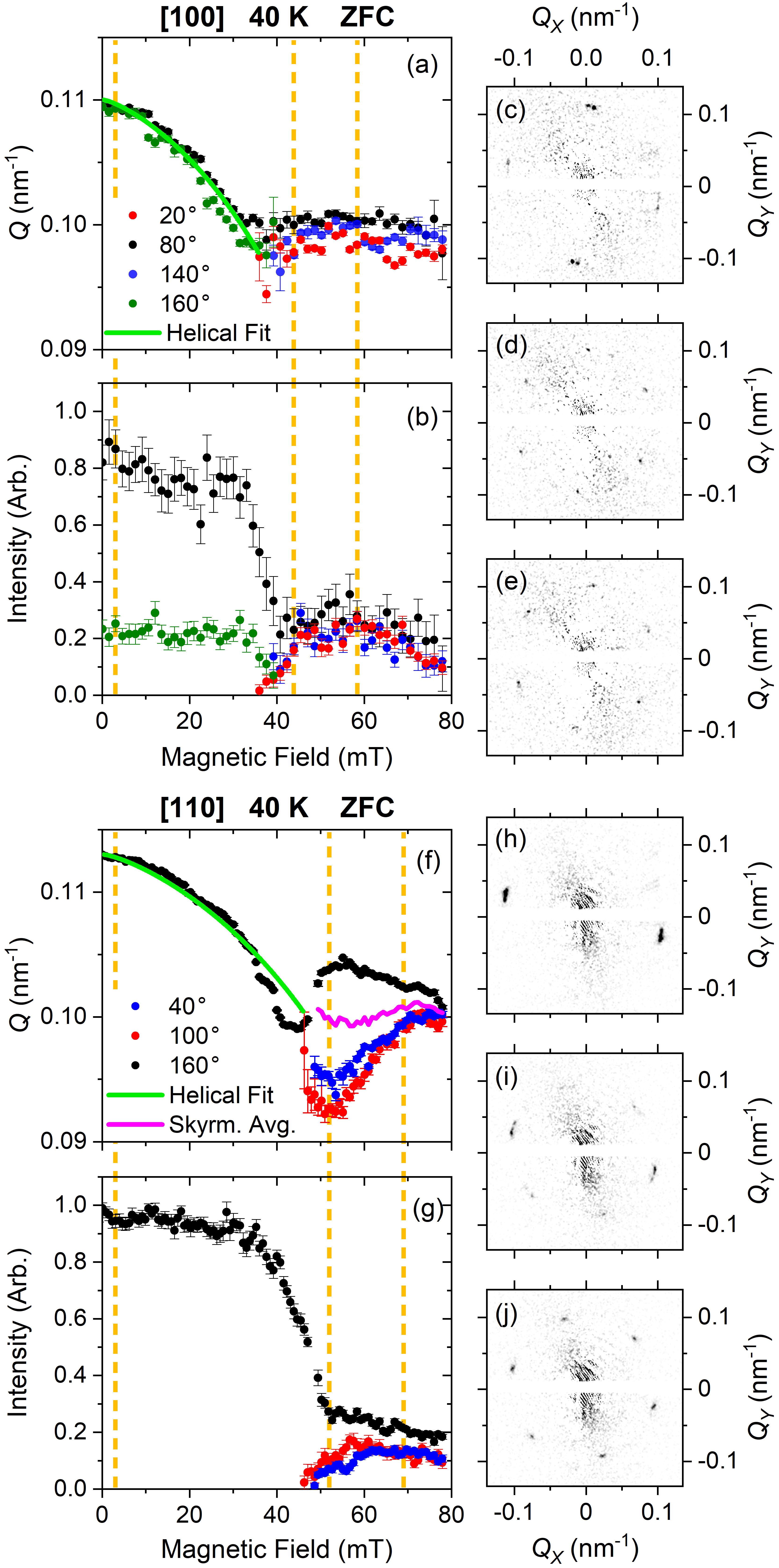}
\caption{Field scan after ZFC to 40~K. (a) and (b) show the field dependence of the $Q$ vectors and peak intensities for the [100] sample. Colors represent pairs of peaks at 160\degree ~(green), 140\degree ~(blue), 80\degree ~(black), and 20\degree ~(red), where angles correspond to the labels on Fig. \ref{fig:Phase} (d).  (c - e) show SAXS patterns for the [100] sample at (c) 3~mT , (d) 44~mT , and (e) 58~mT fields, as indicated by the gold dashed lines on (a) and (b). (f) and (g) show the field dependence of the $Q$ vectors and peak intensities for the [110] sample. Colors represent pairs of peaks at 160\degree ~(black), 100\degree ~(red), and 40\degree ~(blue).  (h - j) show SAXS patterns for the [110] sample at (h) 3~mT, (i) 52~mT, and (j) 69~mT fields, as indicated by the gold dashed lines on (f) and (g). Green lines on (a) and (f) show expected $Q$ dependence of the helical state, while the magenta line on (f) shows the weighted average of the $Q$ values in the skyrmion state.}
\label{fig:40K}
\end{figure}

To investigate the magnetic phases  in more detail, Fig. \ref{fig:40K} shows data for zero field cooled field scans of both samples at 40~K. These plots show significant differences in how the phases evolve with field within the stability regions marked in Fig. \ref{fig:Phase}. Fig. \ref{fig:40K} (a) shows that in the [100] sample all of the pairs of peaks share the same $Q$ value at a given magnetic field. This demonstrates that at low field when two helical domains coexist, they share the same modulation length, suggesting that they arise from the same energy scales, and that at higher fields, the skyrmion lattice is characterized by a symmetric sixfold lattice. Fig. \ref{fig:40K} (b) also supports this, showing that the intensities of the three peaks in the skyrmion state are consistent with one another, while the intensities of the two helical peaks are different, likely arising from different volume fractions of the two domains being sampled in our measurements of a limited 7$\mu$m$^2$ area. 

By contrast, Fig. \ref{fig:40K} (f) shows significantly different behavior for the $Q$ values of the peaks in the [110] sample. In this case, the $Q$ value of the single domain helical state exhibits a distinct minima at the helical-skyrmion transition, and afterwards rises to a substantially higher value than that of the other two peaks. This difference indicates that the skyrmion lattice in this case is distorted, being somewhat shortened along the direction in which the helical state forms ([001] direction). At higher fields, the difference between the $Q$ values of the peaks decreases, indicating reduced distortion. Similarly, the intensities shown in Fig. \ref{fig:40K} (g) for the [110] sample are not the same between the different peaks, consistent with the picture of a distorted skyrmion state. 

For both samples, we can fit the low field $Q$ dependence of the diffraction peaks using the expected model of the distortion of the helical state into a chiral soliton lattice with applied field. Following references \cite{Dzyaloshinskii1965} and \cite{Izyumov1984}, we can write,
\begin{equation}
\frac{\text{E}(\varkappa)}{\varkappa} = \sqrt{\frac{1}{h}},
\end{equation}
where $\text{E}(\varkappa)$ is the complete elliptical integral of the second kind, $h$ is the ratio between the applied field and the critical field ($H / H_c$), and $\varkappa$ is the modulus of the elliptical function, for which we want to solve. Using the value of $\varkappa$ determined by minimizing this equation, the wavelength ($\lambda$) of the distorted helix is,
\begin{equation}
\lambda = \frac{\alpha \varkappa \text{K}(\varkappa)}{\sqrt{h}},
\label{eq:lambda}
\end{equation}
where $\alpha$ is a scale factor, and $\text{K}(\varkappa)$ is the complete elliptical integral of the first kind. Then, $Q = 2\pi / \lambda$, using Eq. \ref{eq:lambda} for $\lambda$. This expression should then be corrected for the demagnetizing field of the sample, for which the magnetic field is expressed as $H = H_a -  N M$, where $H_a$ is the applied field, $M$ is the magnetization, and $N$ is the demagnetization factor, which we set to one for our plate-like samples. The magnetization for use in this expression is theoretically determined by computing the magnetization profile of the distorted helicoid state for $\vec{Q} \parallel \hat{x}$ and $\vec{H_a} \parallel \hat{z}$ as,
\begin{equation}
\begin{split}
\vec{M} & = M_s(0,M_y,M_z), \\
M_y & = \sin(2\text{am}(\sqrt{H}\frac{x}{\varkappa},\varkappa)), \\
M_z & = \cos(2\text{am}(\sqrt{H}\frac{x}{\varkappa},\varkappa)),
\end{split}
\end{equation}
where $\text{am}(u,v)$ is the Jacobi amplitude function, and $M_s$ is the saturation magnetization. This can be conveniently expressed using the Jacobi Elliptic functions $\text{sn}(u,v) = \sin(\text{am}(u,v))$ and $\text{cn}(u,v) = \cos(\text{am}(u,v))$ as,

\begin{equation}
\begin{split}
M_y & = 2 \text{cn}(\sqrt{H}\frac{x}{\varkappa},\varkappa)\text{sn}(\sqrt{H}\frac{x}{\varkappa},\varkappa), \\
M_z & = \text{cn}^2(\sqrt{H}\frac{x}{\varkappa},\varkappa) - \text{sn}^2(\sqrt{H}\frac{x}{\varkappa},\varkappa).
\label{eq:magprof}
\end{split}
\end{equation}

From Eq. \ref{eq:magprof}, the net magnetization can be determined by integrating $M_z$ over one period along $x$, using the previously experimentally determined saturation magnetization of Cu$_2$OSeO$_3$ of 1240~G (0.50 $\mu_B$ / Cu) \cite{Kohn1977}. This is then used in the demagnetizing field equation to arrive at a final expression for $M(H_a)$.

The result of these calculations is shown as the green lines in Fig. \ref{fig:40K} (a) and (f), where $\alpha \propto 1/Q(0)$ and $H_c$ are adjustable parameters set to $H_c = 72$~mT, $Q(0) = 0.110$~nm$^{-1}$ for the [100] sample, and $H_c = 93$~mT, $Q(0) = 0.113$~nm$^{-1}$ for the [110] sample. This curve shows good agreement for the [100] sample from 0~mT all the way up to the transition to the skyrmion state at 35~mT, for both helical domains. For the [110] sample, there is also good agreement from 0 to 35~mT, where the intensity of the helical peak begins to sharply decrease. However, the transition to the skyrmion state indicated by the appearance of the other pairs of two diffraction peaks seems to occur at a higher field of 46~mT. This suggests that the [110] sample likely goes through a broad transition between 35 and 46~mT where there is some mixed state that is not well identified as either purely skyrmion or helical, while the [100] sample transitions more abruptly between states.

Above the helical transition, the $Q$ value in the [100] sample remains approximately constant with field as the skyrmion state appears and vanishes, with a value corresponding to skyrmion spacing of $63.5 \pm 0.5$~nm. For the [110] sample, the three pairs of diffraction peaks each exhibit quite different behaviors with field. However, if we define an average of the $Q$ values weighted by their respective peak intensities, the result, shown as the magenta line on Fig. \ref{fig:40K} (f), also is approximately constant as a function of field, with a value corresponding to skyrmion spacing of $62.5 \pm 0.5$~nm. This indicates that there is no strong energetic preference for the lattice parameter of the skyrmion state to change size with field, inconsistent with previous reports \cite{Okamura2017}. A possible explanation for this discrepancy is that the skyrmion size is more strongly pinned to a certain value in our thinner samples.

\begin{figure}[t]
\includegraphics[width=\columnwidth]{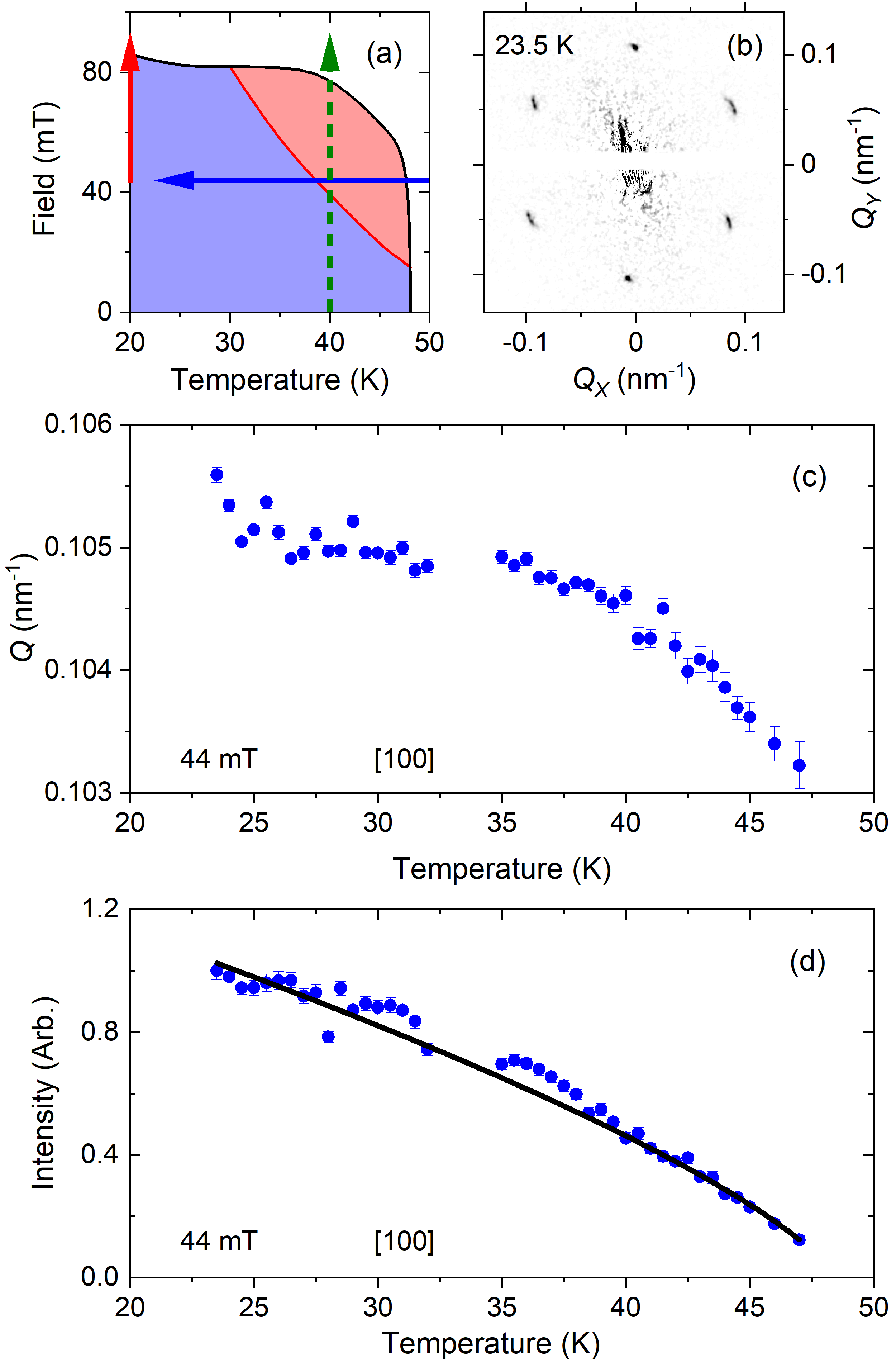}
\caption{Temperature scan of the [100] sample taken while cooling at a fixed field of 44 mT. (a) shows the cooling procedure, (b) shows an example SAXS pattern collected after cooling to 23.5~K, (c) shows the temperature dependence of the $Q$ vector of the skyrmion lattice, and (d) shows the temperature dependence of the scattered intensity. The black line in (d) is a power law fit with a 3D Heisenberg critical exponent of 0.73.}
\label{fig:TScan}
\end{figure}

After measuring the equilibrium phase diagram of the sample using zero field cooled scans, we investigated the metastability of the skyrmions by performing field cooled measurements through the skyrmion state, following the blue arrow on the schematic in Fig. \ref{fig:TScan} (a). We found that for both samples we could create metastable skyrmions at low temperature when cooling through the equilibrium skyrmion region in an applied field. Figure \ref{fig:TScan} (b) shows the scattering pattern measured on the [100] sample at 23.5~K after field cooling in 44~mT. This clearly shows the six-fold symmetric scattering peaks characteristic of the skyrmion lattice, well outside the temperature range where it is seen in the ZFC measurements. 

Figures \ref{fig:TScan} (c) and (d) show the $Q$ value and intensity of the skyrmion peaks as a function of temperature while cooling the [100] sample slowly from 60~K down to 23.5~K  in 44~mT. These both show a smooth increase with decreasing temperature, and no significant anomaly when crossing out of the equilibrium skyrmion pocket at approximately 40~K. This suggests that the metastable skyrmion state carries on smoothly from the equilibrium state, rather than there being an observable phase transition between the two. The intensity data fits to power law behavior with a critical exponent of 0.73. This is the value expected of a three dimensional Heisenberg system  \cite{Holm1993} (which Cu$_2$OSeO$_3$ has been proposed to be  \cite{Zivkovic2014}), if the scattered intensity scales with the square of the magnetization ($I \propto M^2$) as is usually expected \cite{Lovesey1996}.

\begin{figure}[t]
\includegraphics[width=\columnwidth]{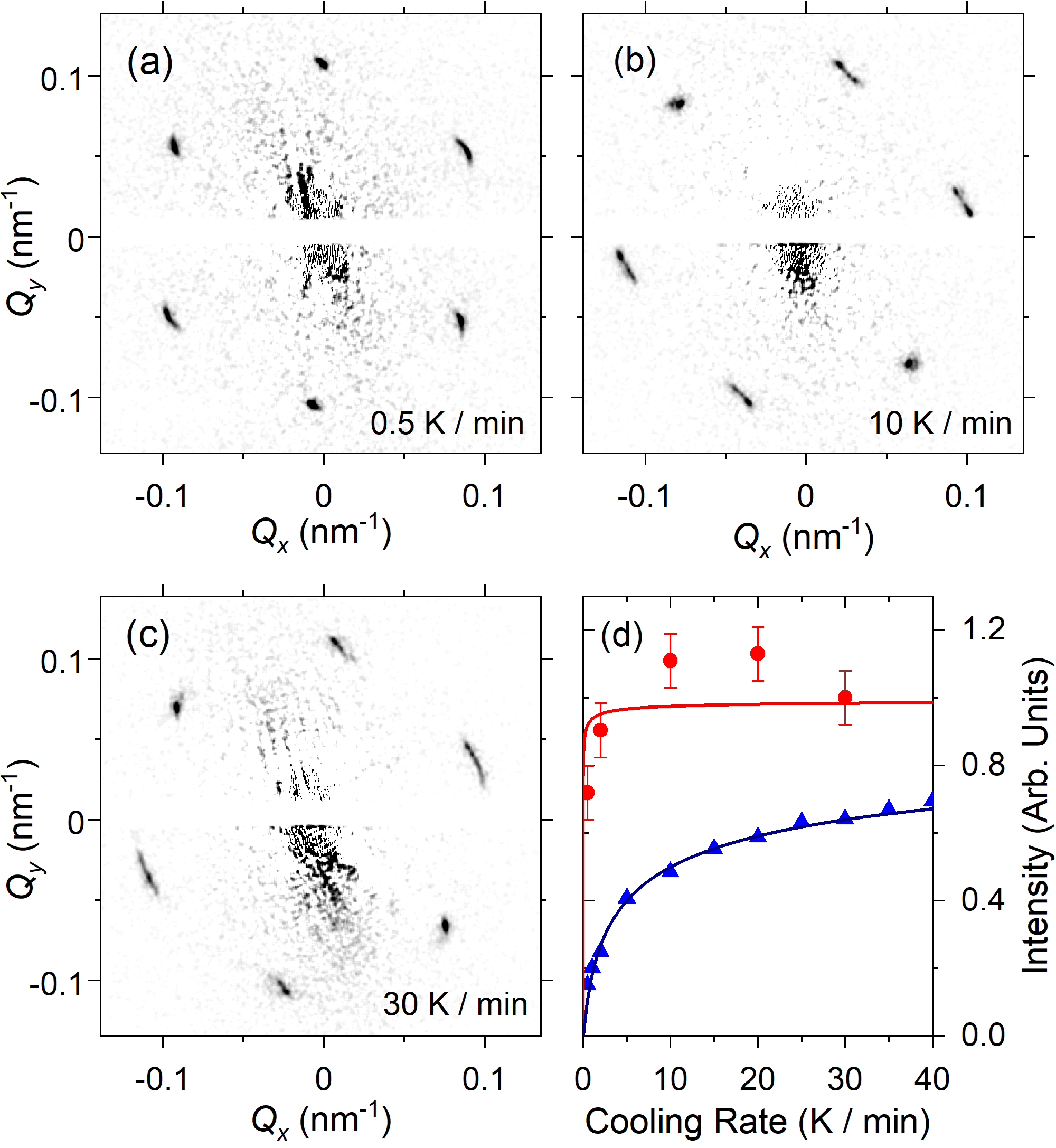}
\caption{Field cooled metastable skyrmion scattering patterns in a [100] orientated Cu$_2$OSeO$_3$ sample measured at 20~K and 28~mT, for cooling rates of (a) $<$~0.5~K/min (22.5 K), (b) 10K/min, and (c) 30K/min. (d) shows the scattered intensity from the skyrmion lattice from each of panels (a-c) in red, and the skyrmion intensity as a function of cooling rate for bulk Cu$_2$OSeO$_3$ in blue \cite{Birch2019}. The navy line shows a fit to the bulk intensity vs. cooling rate data using Eq. A11 from Ref. \cite{Birch2019}, and the red line shows this fit with an adjusted initial temperature of $T_C - 5$~K, appropriate for the thin lamella. Error bars for the blue points are taken as the standard deviation of data in Fig. \ref{fig:Multiple} (c).}
\label{fig:Rate}
\end{figure}

In bulk samples, it has been seen that the population of metastable skyrmions depends strongly on the rate at which the sample is cooled, because of the metastable skyrmions decaying very quickly at high temperature just below the bottom border of the skyrmion pocket \cite{Birch2019}. We therefore investigated this for our thin sample. Figure \ref{fig:Rate} shows SAXS patterns measured at 20~K in 28~mT after field cooling from 60~K for cooling rates of (a) $< 0.5$~K/min, (b) 10~K/min, and (c) 30~K/min. These all show six-fold skyrmion diffraction patterns of similar intensity. Figure \ref{fig:Rate} (d) shows the intensity of the skyrmion scattering peaks as a function of cooling rate (red points), along with the skyrmion intensity as a function of cooling rate for a bulk sample of Cu$_2$OSeO$_3$ from AC susceptibility (blue points), both of which can be taken as proportional to the skyrmion volume in the sample. This shows that while in bulk samples the skyrmion volume raises by more than a factor 4 between a cooling rate of 0.5~K and 30~K, the change in our thin sample over the same range is only 28\% .

The lack of sensitivity of the skyrmion intensity on cooling rate strongly suggests that the thin lamellae do not pass through a region where these metastable skyrmions decay with a very short lifetime. In bulk samples with fields parallel to the [100] direction, the metastable skyrmion lifetime is substantially shorter than 1 minute for temperature 1~K below the skyrmion pocket \cite{Birch2019}. Such a short lifetime would cause a dramatic loss of skyrmion intensity for the slow cooling rate of $<$~0.5~K / min, which is not seen in our thin sample. In addition, the detailed temperature scan in Fig. \ref{fig:TScan} shows no temperature at which there is an observed rapid decay of skyrmions (which would show up as a strong decrease in skyrmion intensity). This demonstrates that the metastable skyrmions are much more stable in thin samples than they are in bulk. 

Enhanced stability of metastable skyrmions in thin lamellae contrasts with theoretical expectations that skyrmions have a lower energy barrier to decay through interfaces than in the bulk of a sample \cite{Cortes2017}, which would result in faster decay in thin samples. Our observations therefore may arise just from the expansion of the equilibrium skyrmion pocket to lower temperature compared to bulk samples. This would result in less thermal energy available for the system to hop over the energy barrier between the skyrmion state and the neighboring magnetic states. Data on bulk samples shows that the skyrmion lifetime increases by an order of magnitude for every 1~K decrease in temperature \cite{Birch2019}. In bulk, the bottom of the skyrmion pocket is 3~K below $T_C$; by comparison we see that it is about 5~K below $T_C$ at 28~mT in our [100] lamella. We therefore might expect the lifetimes to be $100$ times longer than those in bulk at a similar distance below the skyrmion pocket, which would be too long to observe within the time-frame of our experiments. 

To better illustrate the expected difference in behavior when just the temperature of the skyrmion boundary is changed, we fit the intensity vs. cooling rate data for the bulk sample in Fig. \ref{fig:Rate} (d) with equation A11 from Ref. \cite{Birch2019}, and show the result as a navy line. Using the fit values, but changing the start temperature of the cooling process to the $T_C - 5$~K appropriate for the thin lamella, we get the red line shown in Fig. \ref{fig:Rate} (d). This is nearly flat, consistent with our data, with a rapid upturn only for very small cooling rates. This therefore shows that the dramatic difference in the metastable decay behavior may just be explained by the temperature offset, without requiring any specifics of the energetics of the skyrmion state to be changed. 

\begin{figure}[t]
\includegraphics[width=\columnwidth]{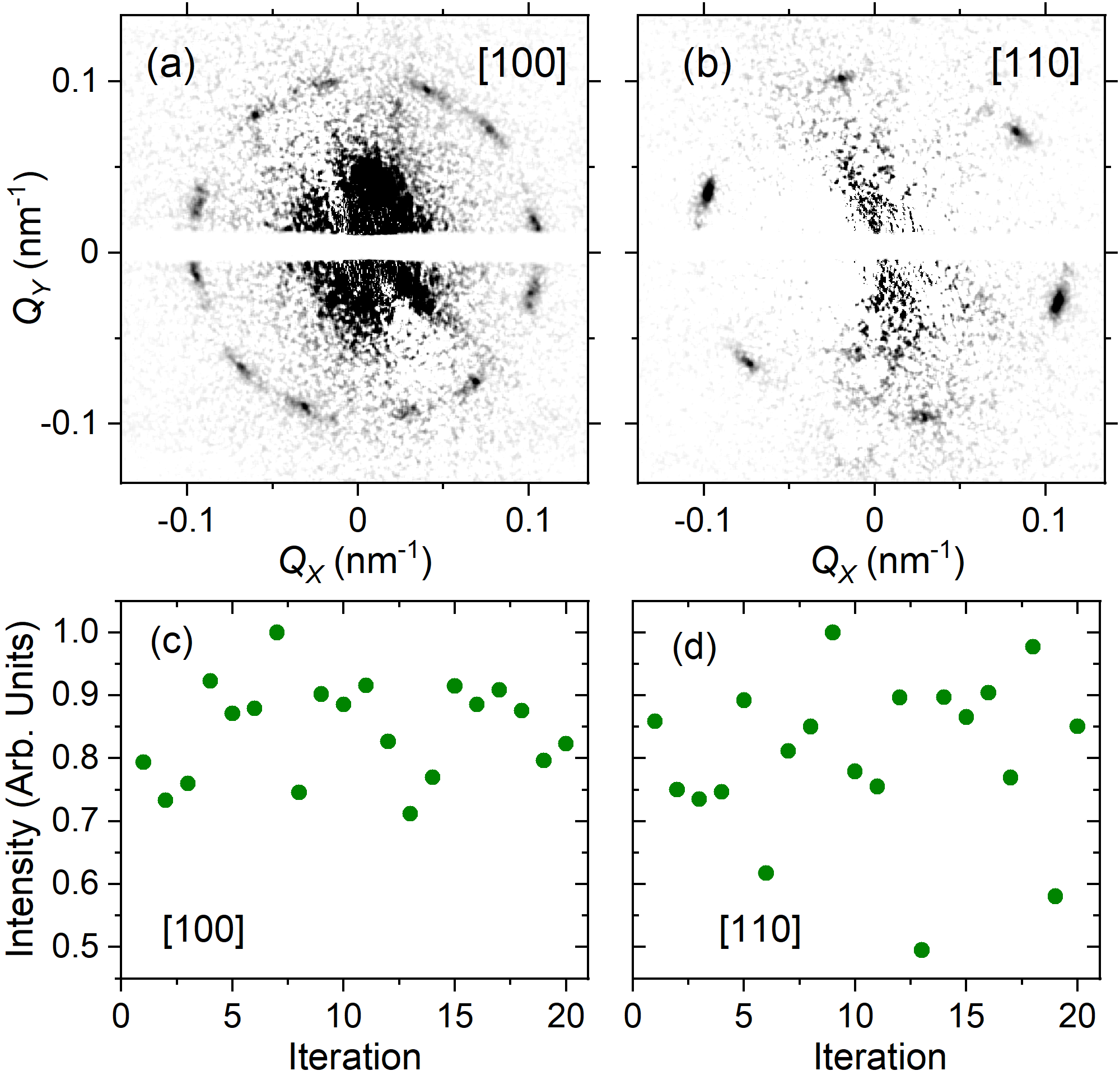}
\caption{Sum of images captured for 20 different field cools from 60K to 20K at 30 K/min in 54~mT for (a) [100] sample and (b) [110] sample. Intensity of the skyrmion peaks for each measurement normalized to the highest intensity run is shown in (c) for the [100] sample and (d) for the [110] sample.}
\label{fig:Multiple}
\end{figure}

During this experiment, we noted that the exact skyrmion state varied significantly between different field cools. To better characterize this, we performed 20 consecutive field cools from 60~K to 20~K at 30~K/min in 54~mT on each sample, measuring a diffraction pattern at 20~K after each cool down. The sum of these patterns is shown in Fig. \ref{fig:Multiple} (a) for the [100] sample and \ref{fig:Multiple} (b) for the [110] sample. These clearly show that there is a strong preference for a single skyrmion domain to form in the [110] sample, while there are two separate skyrmion domains that are equally likely to form in the [100] sample, giving the 12 diffraction spots in the summed image. These two domains correspond to skyrmions forming with a pair of diffraction spots aligned with one of the two different helical domains we see at zero field, and hence the domains are aligned at angles consistent with the angle between the two helical domains. On different field cools on the [100] sample we found that we could observe either of these domains, or both, in identical conditions. The exact formation of the skyrmion domain is therefore somewhat random. The [110] sample also showed some variation, but only with small changes in the angle of the skyrmion domains centered around a single preferred position, giving the slight smearing of the diffraction peaks seen in Fig. \ref{fig:Multiple} (b).

Figures \ref{fig:Multiple} (c) and (d) show the total skyrmion peak intensity for each of the field cools. The 32\% variation between the most and least intense in the [100] sample is comparable to the 28\% change seen between the 30~K/min and 0.5~K/min measurement in Fig. \ref{fig:Rate}. It therefore seems likely that a significant part of the apparent change in skyrmion intensity with cooling rate in Fig. \ref{fig:Rate} comes from random variations in peak intensity between different field cool iterations. We have visually represented this on Fig. \ref{fig:Rate} (d) by showing error bars corresponding to the standard deviation of the data in Fig. \ref{fig:Multiple} (a). This would also explain the non-monotonic change in intensity with cooling rate in Fig. \ref{fig:Rate} (d), where the 30~K/min data is 12\% lower than the 20~K/min. Overall, this reinforces the picture that cooling rate does not have any significant effect on the metastable skyrmion population.

\begin{figure}[t]
\includegraphics[width=\columnwidth]{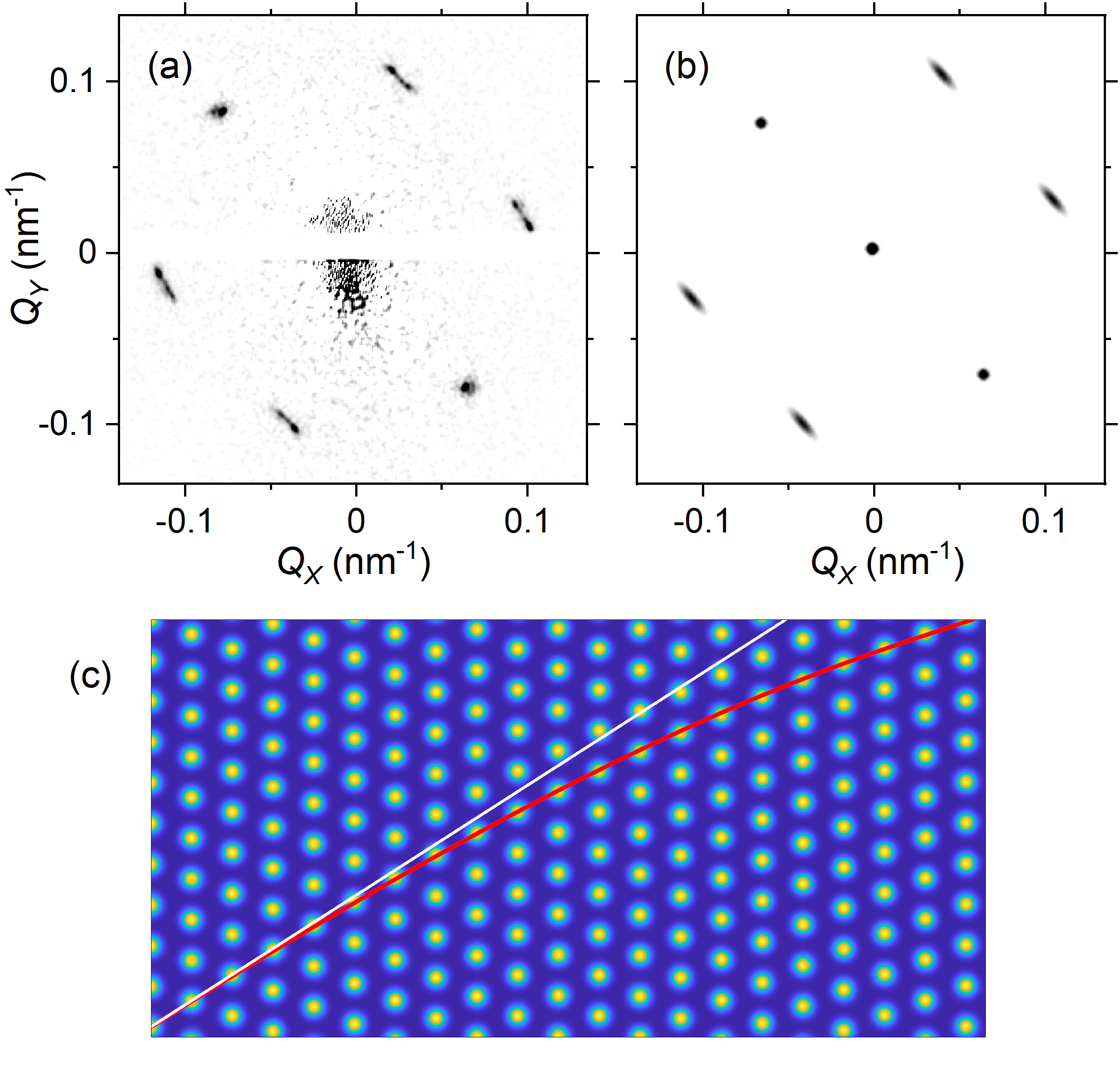}
\caption{(a) SAXS pattern of the [100] sample collected at 20K after field cooling at 10~K/min in 28~mT. (b) Fourier transform of the distorted skyrmion lattice shown below. (c) Schematic of a distorted skyrmion state with shear defects along one direction. The lines are guides to the eye: red line traces a line of skyrmions in the distorted state, and the white line is a reference straight line where an undistorted lattice would be.}
\label{fig:Streak}
\end{figure}

Some of the low temperature skyrmion SAXS images we observed, such as the 10~K/min FC data in Fig. \ref{fig:Rate} (b), show significant broadening of two of the pairs of peaks, but not the third pair. This can be explained by a varied shear distortion of the skyrmion lattice along one direction, resulting in lines of skyrmions that are displaced with respect to the expected perfect 60 degree hexagonal lattice positions. Figure \ref{fig:Streak} (c) shows a model of a possible distortion, which yields the Fourier transform shown in \ref{fig:Streak} (b). This sort of distortion qualitatively matches the observed scattering pattern in Fig \ref{fig:Streak} (a), and can be explained by shear forces caused by a spatially non-uniform magnetic field \cite{Brearton2019}. A small degree of such spatial non-uniformity in our magnetic field across the sample would be expected from demagnetizing field effects including from slight variations in the sample thickness (expected $<10$~nm differences), and because the field is applied by four permanent magnets arrayed around the sample, rather than a solenoid that may give a more uniform field.

\begin{figure}[]
\includegraphics[width=\columnwidth]{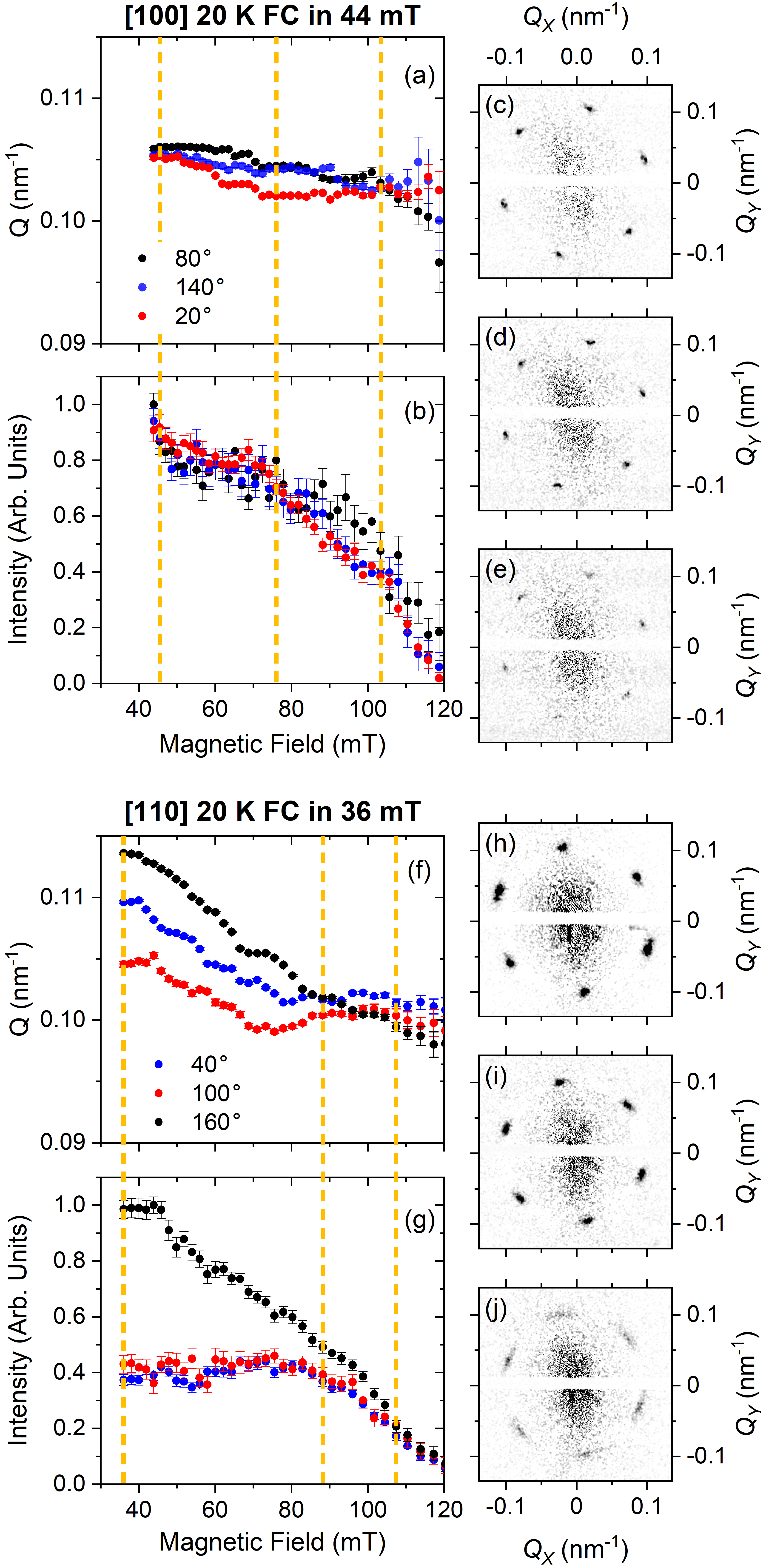}
\caption{Field scan of the metastable skyrmion state after field cooling at 30K/min to 20~K in 44~mT [100] and 36~mT [110].  (a) and (b) show the field dependence of the $Q$ vectors and peak intensities for the [100] sample. Colors represent pairs of peaks at 140\degree ~(blue), 80\degree ~(black), and 20\degree ~(red), where angles correspond to the labels on Fig. \ref{fig:Phase} (d).  (c - e) show SAXS patterns for the [100] sample at 46~mT (c), 76~mT (d), and 103~mT (e), as indicated by the gold dashed lines on (a) and (b). (f) and (g) show the field dependence of the $Q$ vectors and peak intensities for the [110] sample. Colors represent pairs of peaks at 160\degree ~(black), 100\degree ~(red), and 40\degree ~(blue).  (h - j) show SAXS patterns for the [110] sample at 36~mT (h), 88~mT (i), and 107~mT (j) fields, as indicated by the gold dashed lines on (f) and (g).}
\label{fig:20KMS}
\end{figure}

To further characterize the metastable skyrmion state, we performed a field scan at 20~K of both samples after field cooling. Figure \ref{fig:20KMS} (a) and (b) show the $Q$ value and intensity for the skyrmion peaks of the [100] sample as a function of field. The $Q$ value is the same for all three pairs of diffraction peaks, and shows only a very slight decrease with increasing field. This behavior is consistent with what was seen for the field scan of the equilibrium skyrmion state. By contrast, the skyrmion intensity smoothly decreases from its initial value to zero by a magnetic field of 120~mT that is substantially higher than the 87~mT the helical state vanishes at in the zero field cooled scans, or the 82~mT maximum field we saw the skyrmions exist to in equilibrium at higher temperature. This increased field region highlights the high stability of the metastable state once it has been quenched down to low temperature.

 The data for the [110] metastable skyrmion field scan in Fig. \ref{fig:20KMS} (f) and (g) paints a similar picture to the [100] data. The $Q$ dependence is similar to the zero field cooled scan, with the three peak pairs initially forming at different $Q$ values, and the 160 degree peak starting with a higher intensity,  giving the elliptically distorted skyrmion state shown in Fig. \ref{fig:20KMS} (h). As the field increases, $Q$ slightly decreases, and the intensities and $Q$ values eventually become equal, giving a symmetric scattering pattern as shown by \ref{fig:20KMS} (i). The skyrmions in this orientation also vanish at about 120~mT, significantly above the top boundary of the helical state.
 
 For both samples, the highest field states we could measure (Figures \ref{fig:20KMS} (e) and (j)) show azimuthal broadening of the diffraction spots. This indicates substantial rotational disorder appearing in the skyrmion lattice. Annihilation of individual skyrmions throughout the lattice, and subsequent slight space-filling motion of the remaining skyrmions, moving towards a skyrmion liquid state, would explain such rotation disorder. The presence of this disorder only at high fields may suggest that the metastable skyrmions disappear first at the edges of the lattice, and only begin to disappear in the center at high field. This would be consistent with previous LTEM work on skyrmions in FeGe which showed that edge skyrmions are generally less stable \cite{Peng2018}.

\section{Conclusions}

In conclusion, we have investigated the equilibrium and metastable skyrmion lattice in two 200~nm thick Cu$_2$OSeO$_3$ lamellae aligned in the [100] and [110] orientations. We find that while the region of skyrmion stability is similar for the two samples, there are significant differences in how the state evolves with magnetic field, with the [110] sample showing a pronounced asymmetric distortion in the skyrmion state at low field that slowly disappears as the field is increased, while the skyrmion state in the [100] sample is always symmetric. Furthermore, we find that the skyrmion lattice in the [100] sample has two distinct preferred orientations, aligning with the in-plane [010] and [001] directions, while the [110] sample only has one, aligned with the single in-plane [001] direction. Finally, our measurements of the metastable skyrmion state in these samples shows that it is significantly more stable in thin lamella than it is in bulk crystals. This contrasts with previous theoretical expectations of skyrmions decaying through surfaces. However, this might be explained by the lower temperature of the equilibrium skyrmion state boundary and hence lower thermal energy available for the system to hop over energy barriers.

\begin{acknowledgements}
We acknowledge Diamond Light Source for time on Beamline I10 under Proposal SI19421-1. This work was financially supported by the UK Skyrmion Project EPSRC Programme Grant (EP/N032128/1). M.~N.~Wilson acknowledges the support of the Natural Sciences and Engineering Research Council of Canada (NSERC). 
\end{acknowledgements}

\end{document}